\documentclass[aps,prb,twocolumn,superscriptaddress, show pacs]{revtex4-1}
\usepackage{bm, amsmath}
\usepackage{graphicx}
\usepackage{color}
\usepackage{epstopdf}
\usepackage{graphicx}
\usepackage{dcolumn}
\usepackage{bm}
\usepackage{subfigure}

\begin{document}

\title{Skyrmion instabilities and distorted spiral states in a frustrated chiral magnet}

\author{Thomas T. J. Mutter}
\thanks{thomas$\_$mutter@hotmail.com}
\affiliation{Chirality Research Center, Hiroshima University, Higashi-Hiroshima, Hiroshima 739-8526, Japan}
\affiliation{Zernike Institute for Advanced Materials, University of Groningen, Nijenborgh 4, 9747 AG Groningen, The Netherlands}

\author{Andrey O. Leonov}
\thanks{leonov@hiroshima-u.ac.jp}
\affiliation{Chirality Research Center, Hiroshima University, Higashi-Hiroshima, Hiroshima 739-8526, Japan}
\affiliation{Department of Chemistry, Faculty of Science, Hiroshima University Kagamiyama, Higashi Hiroshima, Hiroshima 739-8526, Japan}
\affiliation{IFW Dresden, Postfach 270016, D-01171 Dresden, Germany} 

\author{Katsuya Inoue}
\thanks{kxi@hiroshima-u.ac.jp}
\affiliation{Chirality Research Center, Hiroshima University, Higashi-Hiroshima, Hiroshima 739-8526, Japan}
\affiliation{Department of Chemistry, Faculty of Science, Hiroshima University Kagamiyama, Higashi Hiroshima, Hiroshima 739-8526, Japan}

\date{\today}

\begin{abstract}
{Magnetic skyrmions are particle-like topological excitations that recently generated much interest as candidates for future spintronic devices based on skyrmion small size, enhanced topological stability, and/or mutual interaction.
Here we examine the properties of isolated  skyrmions in a frustrated chiral magnet with competing Dzyaloshinskii-Moriya and frustrated exchange interactions.
We show that the skyrmion size drastically decreases even for small values of competing stabilization mechanisms.
Skyrmion mutual interaction remains attracting as is inherent for frustrated skyrmions, but the value of the Dzyaloshinskii constant regulates the number of minima in the interaction potentials. 
Moreover, the constructed  phase diagrams for a chiral helimagnet contain a distorted spiral state that can be considered as a buffer between the helicoidal and conical one-dimensional modulations.
The  formulated concepts may further enhance the functionalities of spintronic devices. 
In particular, the controlled instability of skyrmions with respect to the conical state allows to obtain bimeron-like structures.
Moreover, our results provide physical insight into the chiral states in the magnetic systems, e.g., in MnSi$_{1-x}$Ge$_x$.
%
%
}
\end{abstract}

         
\maketitle

\begin{figure}
\includegraphics[width=0.99\columnwidth]{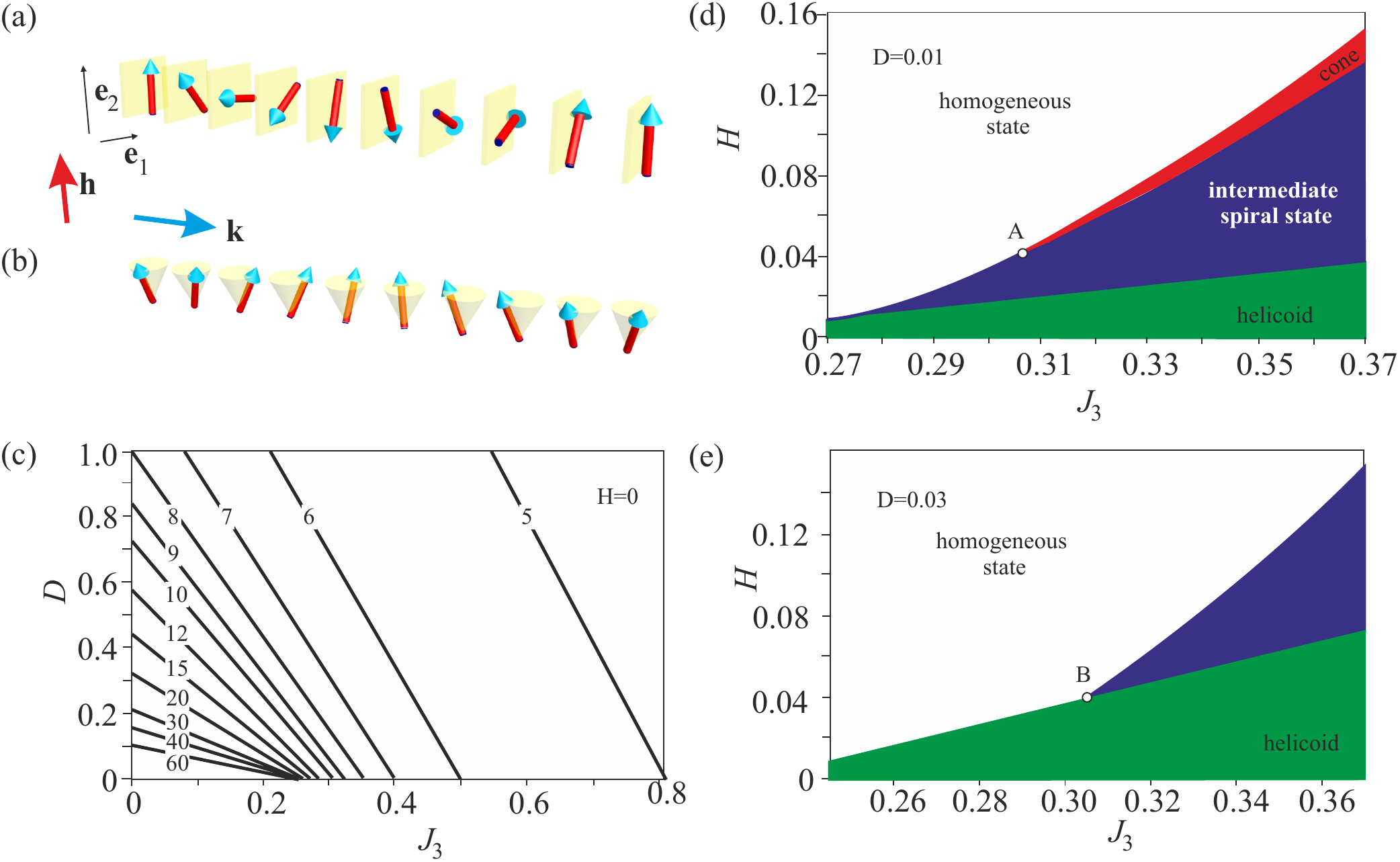}
\caption{
(color online) One-dimensional helicoidal (a) and conical (b) modulations preferred, correspondingly, by DMI and FEI. In a helical
”array” (a) the magnetization rotates in the plane spanned by the orthogonal unity vectors $\mathbf{e}_1$ and $\mathbf{e}_2$ and the rotation sense is determined by the sign of Dzyaloshinskii constant $D$. In the conical state (b), on the contrary the sense of rotation as well as the rotation plane are arbitrary and are not fixed by FEI. (c) Contour plot spanned by $D$ and $J_3$ for zero magnetic field. The contours correspond to constant values of the wave length $\lambda$ (values on the contour lines) and exhibit a drastic change of $\lambda$ for competing DMI and FEI. (d), (e) Magnetic phase diagrams ($J_3$ - $H$) of the 1D solutions for model (\ref{energy}) parametrized by the DMI constant $D$.   Filled areas designate the regions of thermodynamical stability of corresponding modulated phases:  conical phase (red shading); helicoid (green shading); DSS (blue shading). White shading stands for the region of isolated skyrmions. With increased $D$ the regions of cones and DSS are squeezed out into the area of high $J_3$ values. 
\label{PD}
}
\end{figure}

\textit{Introduction. }
%
%
In magnetic compounds lacking inversion symmetry, the Dzyaloshinskii-Moriya interaction (DMI) \cite{Dz,Moriya}  generally described by a spin vector product, $\mathbf{D}_{ij}\cdot(\mathbf{S}_i\times\mathbf{S}_j)$,  plays a crucial role in destabilizing the homogeneous ferromagnetic arrangement and twisting it into a spiral state \cite{Dz1} ($\mathbf{D}_{ij}$ is the Dzyaloshinskii vector). 
Within a continuum approximation for magnetic properties, the DMI is expressed by Lifshitz invariants \cite{LI}, energy terms linear with respect to the first spatial derivatives of the magnetization that arise in certain combinations and depend on crystal symmetry \cite{Bogdanov89}.
At zero magnetic field, helices  are single-harmonic modes \cite{Dz1} $\mathbf{S}=  \mathbf{e}_1 \cos \left(\mathbf{k} \cdot \mathbf{r} \right)+ \mathbf{e}_2 \sin  \left(\mathbf{k}\cdot \mathbf{r} \right)$ with the wavelength $\lambda_D=2\pi/\arctan{(D/J_1)}$ defined by the competing DMI $D$ and the exchange interaction $J_1$ ($\mathbf{e}_1$, $\mathbf{e}_2$,  are the unit vectors in the plane of the magnetization rotation orthogonal to the wave vector $\mathbf{k}$, Fig. \ref{PD} (a)).
%
%
In the magnetic field applied perpendicular to $\mathbf{k}$, the spirals become elliptically distorted and gradually expand ($\lambda_D\rightarrow\infty$) into a system of isolated non-interacting 2$\pi$-domain walls \cite{Bogdanov94,Togawa} (kinks) for the critical field   $h_H=H_H/H_D=\pi^2/4=0.616$, $H_D=D^2/J_1$.
%
%

Another example of spiral stabilization is represented by a chain of Heisenberg spins with ferromagnetic interaction ($J_1>0$) between nearest-neighbor (NN) spins and antiferromagnetic interaction ($J_3<0$) between next-NN spins \cite{frustration}. 
A spiral state appears for $J_3/J_1>1/4$ and retains a single-harmonic character in an applied magnetic field: the spins rotate within a cone surface with the fixed spin component along the applied field, $S_z = (4 H J_3)/(J_1 - 4 J_3)^2$ (Fig. \ref{PD} (b)), with the subsequent  transition   into the homogeneous state. The wave length $\lambda_F=2\pi/\arccos{(J1/4J_3)}$ does not depend on the field.
In the continuum models, the frustrated exchange interactions (FEI) give rise to the energy contributions with higher-order spatial derivatives \cite{frustration,Mostovoy} of the magnetization what makes it similar to the original idea of Tony Skyrme\cite{Skyrme}. 

In the last few years, a renewed interest in non-centrosymmetric and frustrated magnets has been inspired by the discovery of two-dimensional
localized modulations, commonly called magnetic skyrmions \cite{Bogdanov94,yuFeCoSi,Muehlbauer09}.
The role of DMI and/or FEI is to protect such localized particles from radial instability and thus to overcome the constraints of the Hobart-Derrick theorem \cite{Hobart,Derrick}. 
Chiral skyrmions were first spotted in bulk cubic helimagnets such as the itinerant magnets, MnSi \cite{Muehlbauer09} and FeGe \cite{Wilhelm11}.
In these systems skyrmions appear forming hexagonal lattices in a small pocket of the temperature-magnetic field phase diagram just below the transition temperature $T_C$, the so-called A-phase \cite{Kadowaki1982,Muehlbauer09}.
Remarkably, the emergence of a Bloch-type skyrmion state in the frustrated centrosymmetric triangular-lattice magnet Gd$_2$PdSi$_3$ \cite{Kurumaji} was recently reported near the ordering temperature (A-phase of frustrated skyrmions) as also predicted by the theory \cite{Okubo}.
Subsequently, chiral skyrmions (isolated and SkL) have been observed in thin layers of cubic helimagnets (Fe,Co)Si \cite{yuFeCoSi} and FeGe \cite{yuFeGe} in a broad range of temperatures and magnetic fields 
what instigated their reputation as information carriers for future spintronic devices.

%
%

The DMI may also be induced in systems with a broken structural inversion symmetry, e.g.,  at surfaces or interfaces \cite{Bogdanov01,Bode07}. 
Experimentally, the effect of the induced DM interaction on skyrmion and spiral states is well studied in Pd/Fe/Ir(111) bilayers \cite{Romming,Romming2}. 
It is believed that the small skyrmion size in these systems is attributed to the competition between the ferromagnetic exchange, the DM and Zeeman interactions in combination with additional small anisotropic terms \cite{Romming,Romming2,LeonovNJP16}. 
Recent density functional theory (DFT) calculations for Pd/Fe/Ir(111) suggest however that the micromagnetic models must be supplemented by the antiferromagnetic exchange between 2nd and 3rd nearest-neighbors \cite{Dupe}. 
Such effect of FEI leads, in particular, to the drastic enhancement of energy barriers and critical fields of skyrmion collapse as well as skyrmion lifetimes \cite{Gomonay}.
%

In the present manuscript as a first step toward systematic study of an isotropic frustrated chiral magnet, we consider  effects imposed by the DMI on the ground one-dimensional (1D) conical state of a frustrated magnet. 
We 
compute the energies and optimal wavelengths of different types of spin spirals and compare their energies to draw a phase diagram spanned by the antiferromagnetic exchange coupling, the DMI  and the applied field strength. We show that the competition between DMI and FEI can result in short periods and give rise to a novel magnetic phase, that we call the distorted spiral state (DSS).
Moreover, we examine the properties of isolated skyrmions (IS) that exist above the  saturation field.
The spins at the outskirts of frustrated skyrmions and antiskyrmions are known to oscillate with decaying amplitude what leads to their attraction \cite{LM}. 
The skyrmion-skyrmion interaction potentials in this case exhibit plenty of local minima and enable complex cluster formation with varying distance between isolated skyrmions.
The increased DMI value is shown to "erase" the oscillations (as well as local minima of interaction potentials) starting from the outer ones. Still, a significant value of $D$ is needed to completely turn skyrmion attraction into repulsion inherent to chiral skyrmions \cite{leonovPHD,LeonovNJP16,Zhang15}.
We also investigate skyrmion and antiskyrmion instability with respect to the conical and spiral states. 
Moreover, we argue that short period magnetic modulations experimentally observed in some compounds, e.g., in MnSi$_{1-x}$Ge$_x$  \cite{Fujishiro,Tanigaki}, may be attributed to the competing effect of DMI and FEI. 


\textit{Phenomenological model.}
We consider classical spins, $\mathbf{S}_i$, of unit length  on a square lattice in the $xy$-plane with ferromagnetic NN $J_1$ and antiferromagnetic 3rd NN $J_3$ exchange interactions. Additionally, we include the DMI with constant $D$ gradually increased starting from 0. 
In such a form, DMI stabilizes skyrmion and helical states of Bloch type.
The $z$ axis is normal to the lattice plane (the $xy$ plane).
\begin{align}
E=
&-J_1 \sum_{\langle i,j\rangle}\mathbf{S}_i\cdot\mathbf{S}_j+J_3 \sum_{\langle\langle\langle i,j\rangle\rangle\rangle}\mathbf{S}_i\cdot\mathbf{S}_j -H \sum_iS_i^z\nonumber\\
&-D \, \sum_{i}(\mathbf{S}_i \times \mathbf{S}_{i+\hat{x}} \cdot \hat{x} 
+ \mathbf{S}_i \times \mathbf{S}_{i+\hat{y}} \cdot \hat{y}).
\label{energy}
\end{align}
$\langle i,j \rangle$ and $\langle\langle\langle i,j\rangle\rangle\rangle$ denote pairs of NN and 3rd NN spins, respectively, and $J_1,J_3>0$. The third term describes the interaction with the magnetic field parallel to the $z$ axis. 
%
In what follows, energy is measured in units of $J_1 = 1$ and distances -- in units of the lattice constant $a$.
The Hamiltonian (\ref{energy}) is used to compute the energy of different spin configurations which can then be compared to determine the optimal spin configuration for various sets of $J_1$, $J_3$, $D$ and $H$ (see the Supplemental Material for the calculation methods).


\begin{figure}
\includegraphics[width=0.99\columnwidth]{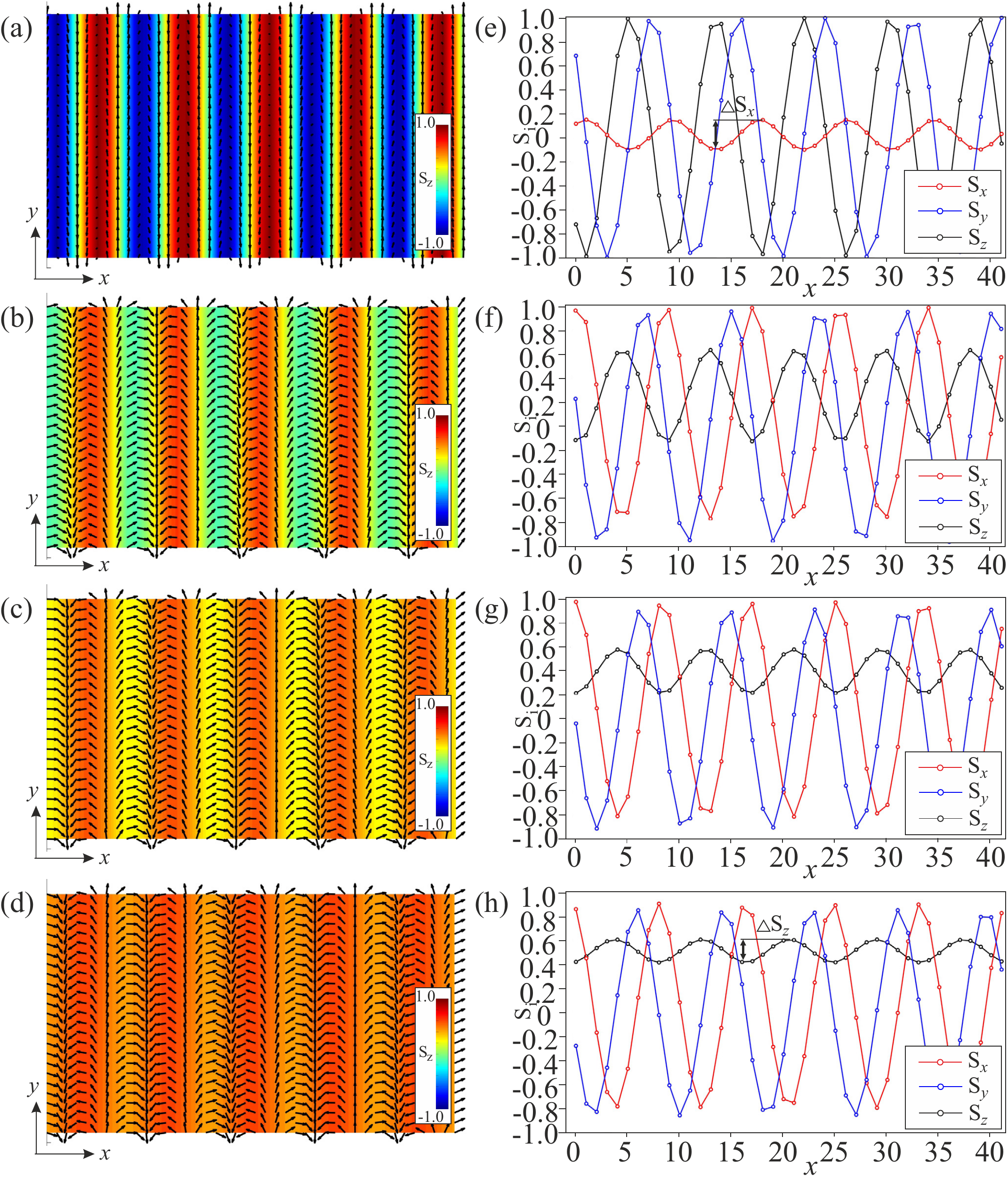}
\caption{
(color online) 
\label{DSS} Real space configurations of DSS for $J_3 = 0.34$, $D=0.05$ and variable magnetic field: $H=0.02$ (a), $H=0.03$ (b), $H=0.04$ (c), $H=0.05$ (d). Color plots in the first column show the out-of-plane spin components, while the arrows -- in-plane ones. The second column shows horizontal linescans of the corresponding color plots. Red, blue and black lines represent $S_x$, $S_y$ and $S_z$, respectively.
DSS is a buffer 1D spiral modulation that gradually develops from the helicoid (a) and transforms into the cone (d).
}
\end{figure}

\textit{1D spiral states.}
%
First, we consider 1D chains of spins with periodic boundary conditions (pbc). 
The number of spins in a chain is varied to address the possible change of an equilibrium spiral period $\lambda$ under an applied magnetic field. 
We prepare a set of spiral states to address the period change with accuracy $0.1$. 
Then, we find the spiral state that minimizes the total energy for each point in a parameter space. 
For example, the number of points 25 in a chain may accommodate different number of wave lengths: one (then $\lambda=25$), two ($\lambda=12.5$) etc. 
By this in particular, we may address the well known process of spiral expansion under an applied magnetic field within the model $J_1 - D$ mentioned in the introduction \cite{Bogdanov94,Togawa} (see the Supplemental Material for details). 
%

The optimal wavelengths of spin spirals for zero field were computed numerically and are plotted in Fig. \ref{PD} (c)  on the plane $D-J_3$  as the straight lines. 
%
%
The slopes of the contours for $\lambda$ can be expressed as $D/J_3$ where $D/J_1 = \tan (2\pi/\lambda)$ and $J_3/J_1 = 1/(4\cos(2\pi/\lambda))$. For fixed $\lambda$, a relation between $D$ and $J_3$ is expressed as
	\begin{equation}
	D = -4\sin (\frac{2\pi}{\lambda})J_3 +  J_1\tan (\frac{2\pi}{\lambda}).
	\end{equation}
Remarkably, the spirals with short periods can be stabilized by either DMI or FEI with relatively large magnitudes.
On the other hand, the same value of the spiral period is achieved for relatively weak competing interactions.
For example,  $\lambda = 9$ can be achieved for $D = 0.18$ ($\lambda_{D} = 34$) and $J_3 = 0.255$ ($\lambda_{F} = 32$). 
In other words, if within the models ($J_1$ - $D$) and ($J_1$ - $J_3$) the spiral states with the same period are stabilized, then within the model ($D$ - $J_1$ - $J_3$) the spiral period would be much smaller, since the interplay between  $D$ and $J_3$ should also be taken into account. 
Moreover, we prove that only one spiral solution is realized differentiating it from the model with competing DMI and  dipole-dipole interaction (DDI) (or FEI and DDI \cite{Hou}), where two energy minima with the properties of spirals and stripe domains (or skyrmions and bubble domains) are known to coexist \cite{Mantel,leonovPHD}.

In an applied magnetic field, as was discussed in the introduction, DMI and FEI  stabilize two different spiral states, correspondingly, helicoids and cones.
In the case of competing DMI and FEI,  we find an intermediate spin spiral phase that bears features of both mentioned 1D modulations. 
In Figs. \ref{PD} (d), (e) we plot "slices" of three-dimensional phase diagrams that contain the stability regions of all three spiral states. 
The phase diagrams are spanned by the exchange coupling strength $J_3$ and the magnetic field for fixed values of DMI $D$.
The low field regime is occupied by the helical spiral, which is the preferred spiral state by the DMI (green shaded region).
In an applied magnetic field, a DSS may appear (blue shaded region), which afterward either undergoes the phase transition with the homogeneous state (white shaded region) or is replaced by the conical phase preferred by FEI (red shaded region). 
%
%
When the DMI strength is sufficiently large, the DSS and the conical phase disappear.

\begin{figure}
\includegraphics[width=0.99\columnwidth]{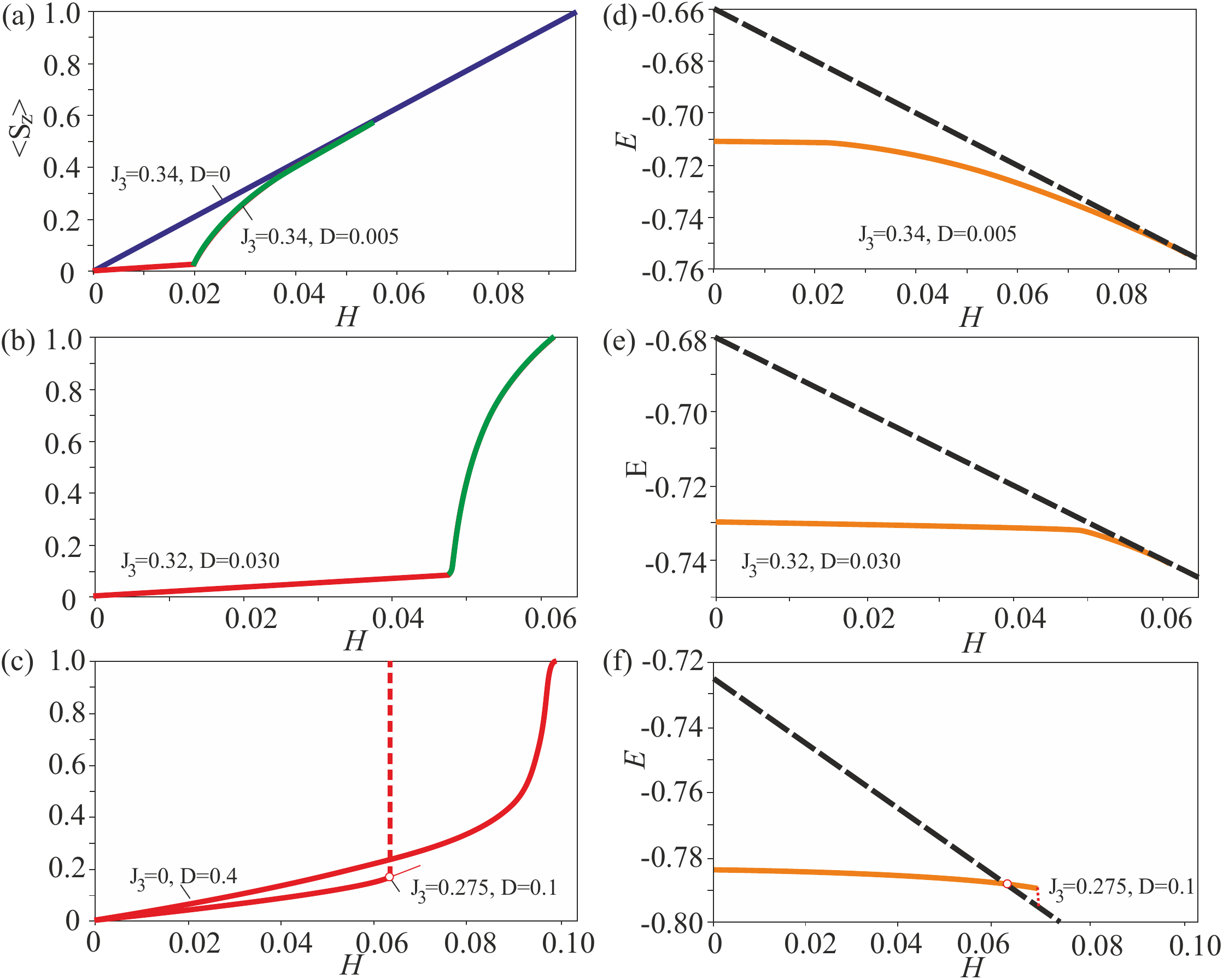}
\caption{
(color online) Representative magnetization curves that indicate all three 1D spiral modulations and possible scenario of magnetization processes that could be identified in the experiments. For conical (blue lines) and helical phases (red lines), magnetization curves represent anhysteretic lines symmetric with respect to the field direction. For $D=0$, the helicoidal magnetization exhibits the drastic increase only near the field of saturation (c). For $D\neq 0$, the helicoids undergo the first-order phase transition  (and are thus accompanied by the hysteresis) 
with the homogeneous state (c). 
Magnetization curves for DSS (green lines) bear pronounced non-linear character 
what is proven additionally by the corresponding energy densities  plotted in the second column (orange curves).
\label{MC}
}
\end{figure}

Fig. \ref{DSS}  shows the real space configurations and variations of the spin components as  functions of $x$ coordinate for the DSS. In contrast to the conical and helical spiral phases, the DSS has all three varying spin components.
In an applied magnetic field, the DSS develops from the helicoid by the second-order phase transition: the rotating magnetization acquires a small oscillating $S_x$-component (Fig. \ref{DSS} (a), (e)). 
In an increased field, DSS transforms into the conical phase with constant $S_z$.
Since some accuracy criteria are involved into the definition of different spiral states, the boundaries of different phases on the phase diagrams may vary. 
The magnetization curves plotted in Fig. \ref{MC} indicate all spiral states and thus can be validated experimentally. Moreover, within the 1D model, the DSS transforms into the homogeneous state by the first-order phase transition. 
In two spatial dimensions, the constructed phase diagrams must be supplemented by the regions of stable skyrmion lattices (SkL).
Within the model ($J_1$ - $D$), the SkLs develop from the helicoids \cite{Bogdanov94} at the field $0.216 H/H_D$ and by the second-order phase transition expand into the homogeneous state for $0.8H/H_D$.

\begin{figure}
\includegraphics[width=0.99\columnwidth]{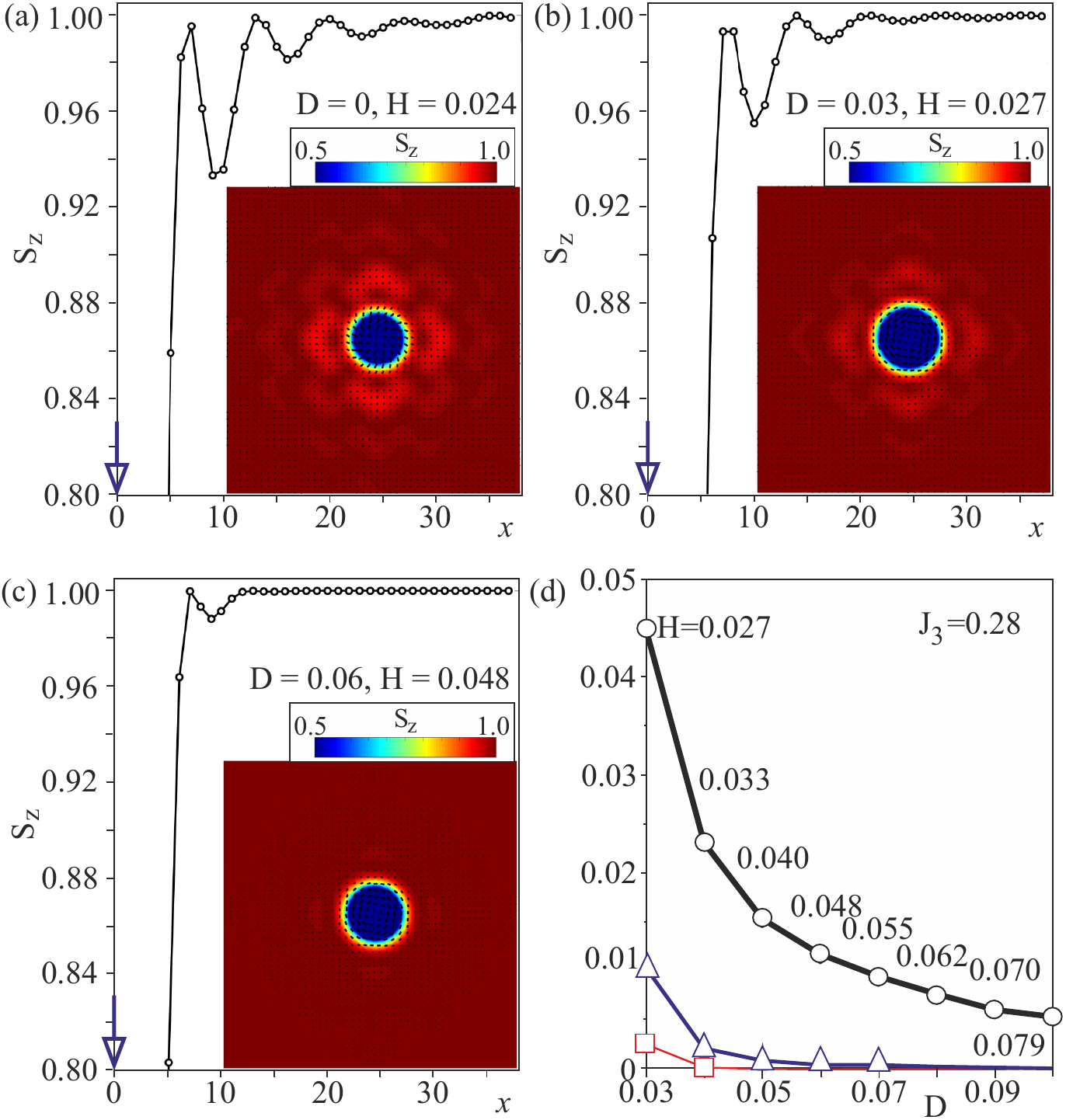}
\caption{
(color online) Fan-like oscillations of the out-of-plane spin components $S_z$ at the ISs outskirts shown for increased value of DMI: $D=0$ (a), $D=0.03$ (b), $D=0.06$ (c). Note that only the range $S_z\in [0.8,1.0]$ is highlighted to make the oscillations identifiable.  Insets show corresponding in-plane components (arrows) and out-of-plane components (color) of spins. The field values in (b), (c) correspond to the inception of ISs' elliptical instability (white circles in Fig. \ref{instability} (h) and spin structures in Fig. \ref{instability} (b), (c)); in (a) -- to the instability with respect to the conical phase (the corresponding spin structure is shown in Fig. \ref{instability} (c)). Amplitudes of first three oscillations are shown in (d) as functions of DMI $D$. Corresponding field values are  indicated near each point.
\label{IS}
}
\end{figure}

\textit{Isolated skyrmions.}
Next, we discuss the properties of isolated skyrmions that are surrounded by the homogeneous state and are thus realized above the saturation fields of cones, helicoids and DSS. 
Frustrated skyrmions ($D=0$ in (\ref{energy})) are known to acquire arbitrary vorticity (i.e., skyrmions and antiskyrmions have the same energy) and helicity (which is a zero mode) \cite{Okubo,LM}. 
Moreover, the spins at the skyrmion outskirt undergo the fan oscillations  (Fig. \ref{IS} (a)) that also give rise to sign changes of the skyrmion-skyrmion interaction potentials and complex cluster formation. 
It is obvious that the additional DMI selects only one type of IS according to the symmetry arguments (Bloch skyrmions in our case) making other  types of ISs metastable particles with higher energy \cite{Dupe2016}. 
Note that Bloch skyrmions are also supported by DDI, since the magnetization component along the radial direction necessarily leads to internal magnetic charges as is observed in Ref. \onlinecite{Kurumaji}. 

The oscillations of spin components with two rotational senses (one of which is not supported by DMI) are also suppressed. 
Remarkably, the extrema of the oscillations do now change their radial positions (Figs. \ref{IS} (b), (c)), but rather become "erased" by the increased DMI (Fig. \ref{IS} (d)).
The global minimum of interaction potential, however, is preserved up to relatively high values of $D$ (comparable with the  magnitude of $J_3$) and weak attraction may be observed. 

\begin{figure*}
\includegraphics[width=1.99\columnwidth]{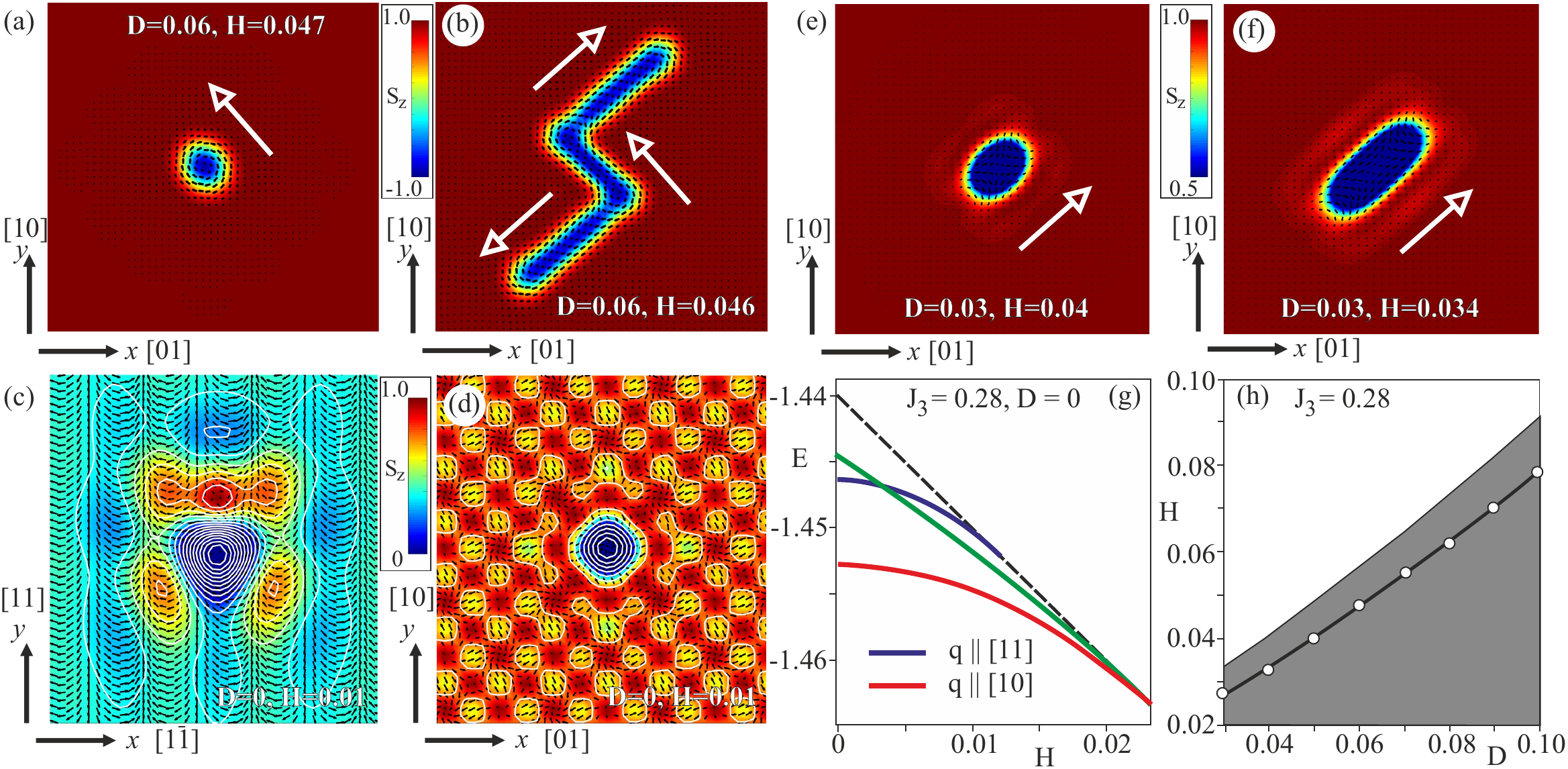}
\caption{
(color online) Instabilities of isolated skyrmions  with respect to the helical (a), (b) and conical (c) 1D spirals as well as with respect to the 2D spin modulation (d). In (a) an IS develops an elliptical distortion along one of the $<11>$ directions. While stretching, it can bend along the complementary $<11>$ (b) and eventually transforms into a spiral domain that occupies the whole space. Elliptical instability of skyrmions occurs slightly below the saturation fields of helicoids (at white circles in (h)).
The $S_z$ spin components are shown as color plots, whereas the black arrows indicate projections of the spins onto the $xy$ plane.The 2D spin structure (d) is realized due to the incompatibility of the conical phase $\mathbf{k}||[11]$ and the square numerical grid with coordinate axes $<10>$. Still, such a spin distribution may appear in nanosystems with confined geometries, since for higher magnetic fields it has lower energy as compared with the cones $\mathbf{k}||[10]$ (green and blue curves, correspondingly, in (g)). To obtain the conical phase $\mathbf{k}||[11]$ with the minimal energy (red curve in (g)), we switch to a new coordinate system with $<11>$ axes. Such a conical phase (c) may accommodate isolated skyrmions with the internal structure resembling bimerons. Antiskyrmions (e), (f), on the contrary, acquire elliptical distortions even above the saturation fields of helicoids (upper boundary of the grey shaded region in (h)). The ratio of ellipticity in this case  is the field-dependent parameter as seen from (e) $H=0.04$ and (f) $H=0.034$. Immediately below the saturation field, antiskyrmions transform into helicoids.
\label{instability}
}
\end{figure*}

With the magnetic field decreased below the saturation value, isolated skyrmions undergo instability with respect to the corresponding 1D spiral state.
The classical example is represented by the tendency of IS to elongate and expand into a band with helicoidal modulations and eventually to fill the whole space (Fig. \ref{instability} (a),(b)) \cite{Bogdanov94,LeonovNJP16}. 
The elongation direction in this case complies with the propagation direction of spirals along $[11]$ inherent to discrete models on the square lattice. 
For lower values of $D$ (e.g., $D=0$, Fig. \ref{instability} (c)), ISs enter the region of cone stability. 
Then in a way similar to bimerons \cite{Murooka,Tretiakov}, their structure acquires regions with the antiskyrmion-like type of the magnetization rotation (red-colored regions) to adjust to the oblique magnetization of the conical spiral. 
Interestingly, pbc may impose an IS instability toward another state with 2D spin arrangement (Fig. \ref{instability} (d)). This state has lower energy than the conical state with $\mathbf{q}||[10]$, but the higher energy as compared with $\mathbf{q}||[11]$ (Fig. \ref{instability} (g)). 
To avoid such an artifact of numerical routines, one should make transition to the  coordinate systems with the axes $x||[1\bar{1}]$ and $y||[11]$ that correctly accommodates  the conical phase with $\mathbf{q}||[11]$ (Fig. \ref{instability} (c)). 

Note that the elliptical instability of IS occurs at the field values lower than the saturation fields of spiral states (white circles in Fig. \ref{instability} (h)).
Antiskyrmions, on the contrary, elongate and fill the space with the spiral modulations as soon as the energy of the spiral state becomes negative as compared with the homogeneous state (in the grey shaded region of Fig. \ref{instability} (h)). 
Interestingly, above the spiral saturation field, antiskyrmions also represent elongated particles, but with the fixed ratio of their ellipticity. Figs. \ref{instability} (e), (f) show metastable antiskyrmions for different values of the field above the region of spiral stability.

\textit{Discussion and Conclusions.}
Recently, short period magnetic modulations have been observed in a series of chiral magnets MnSi$_{1-x}$Ge$_x$ \cite{Fujishiro,Kanazawa} (including MnGe \cite{Tanigaki}) by means of Lorentz transmission electron microscopy and high-field transport measurements. 
Not only the structures with gradual magnetization rotations (like skyrmions or helices) were identified, but also  two distinct three-dimensional hedgehog lattices that incorporate point defects \cite{Fujishiro}. 
Such a short periodicity of magnetic modulations   would require large DMI, which contradicts  however to the band structure calculations \cite{Koretsune} demonstrating smaller values of $D$ with increasing $x$ \cite{Koretsune}. 
Thus, the DMI may not be the primary origin of the short-period helical structure. Instead, the magnetic frustration or Ruderman-Kittel-Kasuya-Yosida (RKKY) interaction \cite{Hayami,Okubo}  can be a possible mechanism as suggested in Ref. \onlinecite{Fujishiro}.

In the present manuscript, we show that the short-size magnetic modulations are induced by the effect of competing DMI and FEI, each with moderate strength values.
Moreover, the constructed phase diagrams exhibit new transitional spiral states and underlie enriched skyrmion properties.
In particular, isolated skyrmions remain attracting also in presence of relatively large DMI. 
The variety of ISs also  extends and includes bimerons (obtained due to IS instability with respect to the cones) and/or elongated antiskyrmions with variable rate of ellipticity.
We stress that the competing DMI and FEI become relevant for skyrmion use  in spintronic devices, since small IS size can be achieved by relatively weak D and J3 strengths. 
Besides, the increased IS stability can result. 
Thus, the proposed effects suggest a new strategy for search of skyrmion hosting materials.


\section{Acknowledgements}

The authors are grateful to Ivan Smalyukh, Katia Pappas, Jun-ichiro Ohe, Istvan Kezsmarki, Hikaru Kawamura and Maxim Mostovoy for useful discussions. This work was funded by JSPS Core-to-Core Program, Advanced Research Networks (Japan) and JSPS Grant-in-Aid for Research Activity Start-up 17H06889. AOL thanks Ulrike Nitzsche for technical assistance.

\end{document}